\documentclass{webofc}

\usepackage[varg]{txfonts}   
\usepackage{hyperref}
\usepackage{subcaption}
\usepackage{url}
\hypersetup{colorlinks=true,citecolor=blue,urlcolor=blue,linkcolor=blue}
\begin{document}
\title{Cross-correlation analysis for cosmic ray flux forecasting}
%
%

\author{\firstname{David} \lastname{Pelosi}\inst{1,2}\thanks{\email{david.pelosi@pg.infn.it}} \and
        \firstname{Fernando} \lastname{Barão}\inst{3} \and
        \firstname{Bruna} \lastname{Bertucci}\inst{1,2} \and
        \firstname{Emanuele} \lastname{Fiandrini}\inst{1,2} \and
        \firstname{Miguel} \lastname{Orcinha}\inst{2} \and
        \firstname{Alejandro} \lastname{Reina Conde}\inst{4,5} \and
        \firstname{Nicola} \lastname{Tomassetti}\inst{1,2} 
}

\institute{Università degli Studi di Perugia, Italy
\and
          INFN – Perugia, Italy
\and
          Laboratório de Instrumentação e Física Experimental de Partículas, LIP - Lisbon, Portugal
\and
           INFN – Bologna, Italy
\and Centro Nacional de Aceleradores (CNA), Sevilla, España
          }

\abstract{
The study presents an effective approach for deriving and utilizing polarity-based cross-correlation functions to forecast Galactic Cosmic Ray (GCR) fluxes based on solar activity proxies. By leveraging a universal correlation framework calibrated with AMS-02 and PAMELA proton flux data under a numerical model, the methodology incorporates Empirical Mode Decomposition (EMD) and a global spline fit. These techniques ensure robust handling of short-term fluctuations and smooth transitions during polarity reversals. The results have significant potential for space weather applications, enabling reliable GCR flux predictions critical for radiation risk assessments and operational planning in space exploration and satellite missions.}
\maketitle
\vspace{-1em}
\section{Introduction}
\label{intro}
The intensity and energy spectrum of GCRs in the heliosphere are strongly influenced by the 11-year solar activity cycle, a phenomenon known as solar modulation. Studying this phenomenon is challenging due to the scarcity of long-term flux records. However, recent space missions, such as PAMELA and AMS-02, have provided valuable time-resolved, multi-channel data on GCR species, enabling significant advancements in solar modulation modeling.
We propose a method to derive universal cross-correlation functions between the modulation parameters of a numerical solution to the Parker transport equation, which governs the propagation of GCRs in the heliosphere, and sunspot number (SSN) as a proxy for solar activity. This work paves the way for developing a fast and practical tool for forecasting fluxes crucial for space weather risk assessment.

\section{Methods}
\begin{figure}[ht]
\centering
\includegraphics[width=12.9cm,clip]{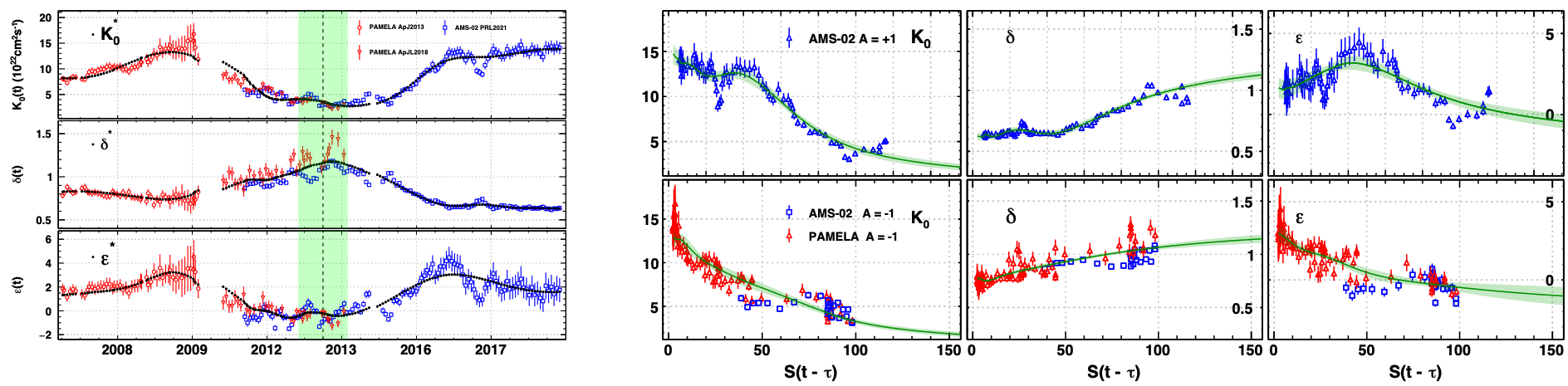}
\caption{\textbf{Left)} Best-fit parameters from proton flux measurements (PAMELA and AMS-02) alongside their EMD-filtered version. The dashed line and shaded region represent the reversal epoch and phase, respectively. \textbf{Right)} Smoothed cross-correlation functions for positive and negative polarity phases, illustrating the evolution patterns of modulation parameters relative to delayed SSN.}
\label{fig:result}       
\end{figure}

The effective model we developed is based on a numerical solution of the Parker equation using the Crank-Nicolson method \cite{Tomassetti2019} under spherical symmetry, with drift effects modeled by an effective convective term $\varepsilon$. The diffusion coefficient is expressed as $K(P,t) = K_0(t) \beta(P) (P/P_0)^{\delta(t)}$, while the species-dependent factor $A/Z$ is embedded in $\beta(P)$. The time-dependent parameters $K_0(t)$, $\delta(t)$, and $\varepsilon(t)$ are fitted using monthly GCR proton flux data from AMS-02 and PAMELA. Short-term variations in the parameter time series, often caused by CMEs or local perturbations not captured by solar proxy, were filtered using the EMD\cite{huang1998} technique. The modes for reconstruction were selected to maximize the Spearman correlation coefficient with respect to the smoothed SSN (S) . Only the modes contributing significantly to the overall correlation were retained, while transient perturbations lacking long-term significance were discarded as illustrated in Fig. \ref{fig:result}. Given the solar wind’s propagation time, a several-month lag exists between solar proxies and variations in GCR measurements, which we addressed using the empirical lag formula from \cite{Tomassetti2022}. We derived cross-correlation functions between the EMD-filtered parameters \(\{K_0^*, \delta^*, \varepsilon^*\}\) and the delayed smoothed SSN. During solar maximum and polarity reversal phases, continuity and smoothness of parameters were maintained using a global spline fit to link the polarity phases with the signed inverse proxy $A/\text{S}(t - \tau)$, where $A = \pm 1 $ represents the polarity. This approach compensates for limited data availability during periods of high SSN.

\vspace{-2em}

\section{Results}
The cross-correlation functions derived, as shown in Fig. \ref{fig:result}, enable the forecasting of modulation parameters and, consequently, GCR fluxes within this numerical framework, using SSN as the sole input. By incorporating the Local Interstellar Spectrum and isotope abundances, flux predictions can be extended into future epochs. Assuming that all species dependence is encapsulated within \(\beta\), this framework is generalizable to forecast any other GCR species. In upcoming publications, we will present detailed results of this approach applied to various nuclei and solar cycles, demonstrating its accuracy in reconstructing GCR fluxes and comparing these with experimental data. Additionally, the model's capability to compute deep space dose rates will be explored, assessing its potential for astronaut radiation risk management.

%
%
%

\end{document}